# Routability in 3D IC Design: Monolithic 3D vs. Skybridge 3D CMOS

Jiajun Shi[1], Mingyu Li[1], Santosh Khasanvis[3], Mostafizur Rahman[2] and Csaba Andras Moritz[1]
[1]Department of Electrical and Computer Engineering, University of Massachusetts, Amherst, MA, USA
[2]School of Computing and Engineering, University of Missouri, Kansas City, MO, USA
[3]BlueRISC Inc.
jiajun@umass.edu, andras@ecs.umass.edu

*Abstract*— Conventional 2D CMOS technology is reaching fundamental scaling limits, and interconnect bottleneck is dominating integrated circuit (IC) power and performance. While 3D IC technologies using Through Silicon Via or Monolithic Inter-layer Via alleviate some of these challenges, they follow a similar layout and routing mindset as 2D CMOS. This is insufficient to address routing requirements in high-density 3D ICs and even causes severe routing congestion at large-scale designs, limiting their benefits and scalability. Skybridge is a recently proposed fine-grained 3D IC fabric relying on vertical nanowires that presents a paradigm shift for scaling, while addressing associated 3D connectivity and manufacturability challenges. Skybridge's core fabric components enable a new 3D IC design approach with vertically-composed logic gates, and provide a greater degree of routing flexibility compared to conventional 2D and 3D ICs leading to much larger benefits and future scalability. In this paper, we present a methodology using relevant metrics to evaluate and quantify the benefits of Skybridge vs. state-of-the-art transistor-level monolithic 3D IC (T-MI) and 2D in terms of routability and its impact on large-scale circuits. This is enabled by a new device-to-system design flow with commercial CAD tools that we developed for large-scale Skybridge IC designs in 16nm node. Evaluation for standard benchmark circuits shows that Skybridge yields up to 1.6x lower routing demand against T-MI with no routing congestion (routing demand to resource ratio < 1) at all metal layers. This 3D routability in conjunction with compact vertical gate design in Skybridge translate into benefits of up to 3x lower power and 11x higher density over 2D CMOS, while TLM-3DIC approach only has up to 22% power saving and 2x density improvement over 2D CMOS.

## I. INTRODUCTION

Migrating to the third dimension for CMOS integrated circuit (IC) design is seen as a way to advance scaling due to fundamental challenges in 2D that stem from device scaling limitations [1], interconnect bottleneck [2] and manufacturing complexities [3]. Conventional approaches such as wafer-to-wafer bonding [4] and monolithic 3D ICs [5] have been extensively explored, and while they show benefits against 2D CMOS they add new design constraints and technology challenges [5]. Among all 3D IC approaches, transistor-level monolithic 3D IC (T-MI) [6] represents the state-of-the-art that uses 3D standard cells for high-density IC design. But it still follows conventional 2D CMOS's routing mindset for inter-cell connections where the standard cells are placed and routed in a two-dimensional plane limiting their accessibility and routability. This in turn causes severe routing congestion [6] in large-scale T-MI ICs diminishing the benefits of this approach and limiting scalability.

Skybridge [7] is a truly fine-grained 3D IC fabric that uses vertically-stacked gates interconnected in 3D on a template of vertical nanowires to yield orders of magnitude benefits over 2D CMOS. Core fabric aspects including device, circuit-style, connectivity [8], thermal management [9] and pathway of manufacturing [10] are co-architected for 3D compatibility. Input/output pins for each vertically-composed gate have multiple points of access both horizontally and vertically which can be reached through architected routing components, as opposed to T-MI which limits pin-access to a 2D plane and relies on conventional routing schemes. Thus Skybridge fully utilizes the vertical dimension providing increased routability to address high-density routing in large-scale ICs.

In this paper, we present a methodology to evaluate and quantify the impact of increased routability in Skybridge on large-scale ICs and compare it with T-MI using benchmark circuits. We develop a novel device-to-system Skybridge IC design flow using commercial CAD tools for large-scale circuit design, encompassing all steps from device characterization, RTL synthesis, cell placement and routing, to system-level density, power and performance evaluation. This enables us to use benchmark circuits such as Data Encryption Standard (DES), low-density parity-check (LDPC) and Joint Photographic Experts Group (JPEG) for evaluation and comparison. We evaluate the routability for each approach using routing demand and routing demand/resource ratio as metrics [11] for benchmark circuits. Evaluation results show that Skybridge achieves up to 1.5x lower routing demand compared to T-MI IC approach. Furthermore, Skybridge designs have no routing congestion even in the interconnect-intensive LDPC circuit, indicated by a maximum routing demand/resource ratio of 0.8 (M5 layer), while T-MI shows routing congestion in LDPC design with a maximum routing demand/resource ratio of 1.25 (M2 layer). Higher degree of routability and compact vertical gate design leads to significantly more wire length saving, power efficiency and density improvement over 2D CMOS than the T-MI approach; Skybridge has up to 3.3x shorter routing wirelength, 3x lower power and 11x higher density compared to 2D CMOS while the T-MI shows 1.4x shorter routing wirelength, 1.3x lower power and 2x density improvement for benchmark circuits.

The rest of the paper is organized as follows: Section II core structures for 3D logic gates and interconnection in Skybridge. Section III in an overview of the CAD tools based device-to-system design flow for Skybridge 3D IC. Section IV shows the routability evaluation and comparison for all technologies. Section V presents the benchmarking results of DES, LDPC and JPEG cores.

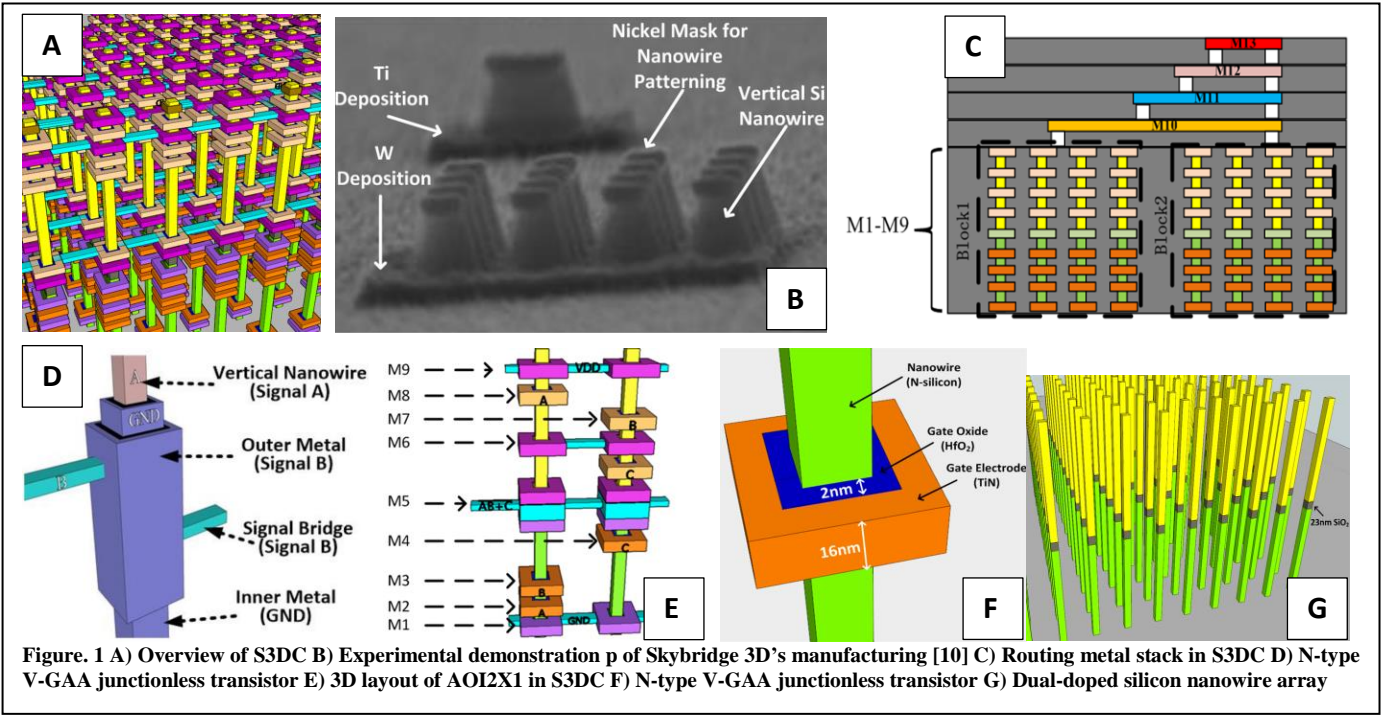

Figure. 1 A) Overview of S3DC B) Experimental demonstration p of Skybridge 3D's manufacturing [10] C) Routing metal stack in S3DC D) N-type V-GAA junctionless transistor E) 3D layout of AOI2X1 in S3DC F) N-type V-GAA junctionless transistor G) Dual-doped silicon nanowire array

## II. OVERVIEW OF SKYBRIDGE

Skybridge offers a family of circuit-styles to be implemented such as NP-Dynamic Skybridge (NP-D-SB) [12] and Skybridge-3D-CMOS (S3DC) [13]. In this work, we focus on S3DC because it uses a static circuit style and fits well into the commercial CAD tools based ASIC design flow. Skybridge-3D-CMOS (S3DC) is a truly fine-grained 3D integration [14], designed with a 3D fabric-centric mindset and providing an integrated solution for all core technology challenges. It expands the fundamental concepts original to Skybridge [7] while realizing a vertically-integrated CMOS circuit style for the first time. Fig. 1A shows the S3DC template; it is built with a regular array of uniform vertical dual-doped nanowires which have p-type doped silicon on the top half and n-type doped silicon on the bot-half (See Fig. 1G). All active components/structures described in this work rely on multi-layer material deposition techniques which have lower cost, and can be controlled to few Angstrom's precision. Manufacturing pathway for S3DC and experimental demonstrations are discussed in [9]. Fig. 1B shows the experimental demonstration of the main manufacturing steps.

Core components including n-type and p-type Vertical Gate-All-Around (n-VGAA and p-VGAA) junctionless transistors [15], are stacked on n-type doped and p-type doped regions of each nanowire to implement complementary logics of static-logic gates. Fig. 1F shows device structure and selected materials of the n-VGAA junctionless transistor; the p-VGAA junctionless transistor has the same device structure but uses different gate metal for channel control which is detailed in [15]. Additional components are needed to enable signal routing in-between these logic gates with compact 3D interconnection and good routability. We use two key components for routing: bridges and coaxial routing structures (Fig. 1D). Bridges are metal lines used as horizontal routing metal to form links between adjacent vertical nanowires. They can be placed at any height on nanowires, and span the required distance by hopping over intermediate nanowires facilitated by coaxial routing structures. The coaxial routing structure consists of concentric metal shells around nanowires separated by dielectric (See Fig. 1D); both of these are unique for Skybridge and enabled by its vertical integration approach. Fig.1D shows an example: signal A is carried by the vertical nanowire and signal B is routed by Bridges; the coaxial routing structure allows signal B to hop the nanowire and continue its propagation. Coaxial routing is enabled by specially configured material structures for insulating oxide and contact metal. Proper materials are chosen and deposited around nanowires to form low-resistivity interconnection between the silicon and the metallic bridge; Details of the contact structure and resistance evaluation are presented in [10].

The vertically-composed 3D logic gates are implemented through stacking contacts and VGAA transistors on vertical nanowires. Fig. 1E shows the layout of a 3-input 3D AND-OR-INVERTER (AOI2X1) gate that is built by using 2 nanowires and 6 VGAA transistors. In total 9 metal layers (M1-M9) are used in the design of S3DC standard cell (See Fig. 1E): M9 is used to place VDD rails which consist of bridges and bridge-to-nanowire contacts, VSS rails with similar structure are placed in M1, M5 is used to place routing components that form the output port, n-VGAA transistors can be placed in three layers M2-M5 and p-VGAA transistors can be placed in three layers M6-M8. This way, each cell is designed with a maximum fan-in of 3. And the feature sizes of contact metal, bridge, VGAA transistors and the nanowire pitch are designed following the design rules as described in [14]. Uniform design rules are applied for each metal layer, and uniform metal thickness and metal-to-metal spacing for each metal layer are set to form uniform vertical routing grids (See Fig. 1E). Additional metal layers (M10-M13) can be added on the top of nanowires array (See Fig. 1C) for providing necessary routing resources in large-scale designs.

## III. SYSTEM-LEVEL DESIGN AND EVALUATION METHODOLOGY

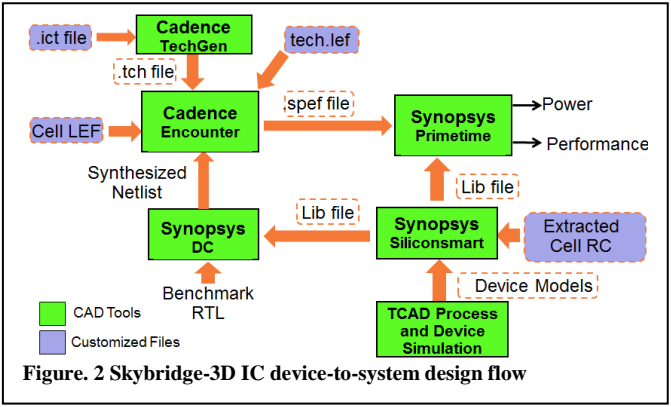

**Figure. 2 Skybridge-3D IC device-to-system design flow**

Fig. 2 shows a device-to-system design flow for mapping the vertical 3D gates based design in S3DC: it mainly includes TCAD based simulations of n- and p-type VGAA junctionless transistors, characterization of standard cell timing and power (Lib file), characterization of interconnect capacitance and resistance table (.tch file), RTL [24] synthesis, placement and route for layout generation, power and performance evaluation. It is a full standard ASIC design flow that is based on commercial CAD tools, creating a solid simulation process.

### A. Device Simulation

VGAA junctionless transistors are used as active devices, and are formed on nanowires through consecutive material deposition steps. VGAA junctionless transistors use uniformly doped with no abrupt variation in Drain/Source/Channel regions that simplifies manufacturing requirements - chosen especially for this fabric. In S3DC, both n- and p-type vertical transistors are VGAA junctionless transistors whose channel conduction is modulated by the workfunction difference between the heavily doped channel and the gate [15]. Titanium Nitride (TiN) and Tungsten Nitride (WN) are chosen for n-type and p-type transistors

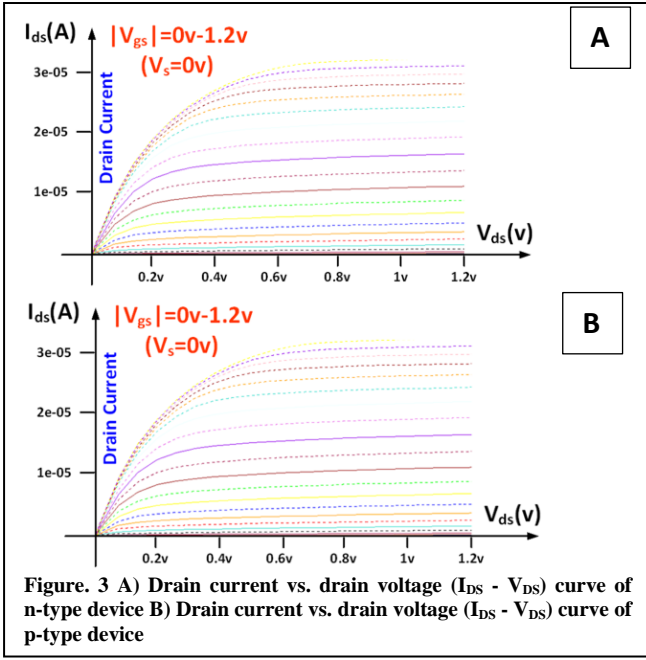

**Figure. 3 A) Drain current vs. drain voltage ($I_{DS} - V_{DS}$) curve of n-type device B) Drain current vs. drain voltage ($I_{DS} - V_{DS}$) curve of p-type device**

respectively to provide proper workfunction [14][15]. 3D TCAD Process and Device simulations [16] are used to extract the device characteristics, shown in Fig. 2. The n-type device had an ON current of 30μA, and OFF current 0.1nA. The p-type device had an ON current of 26μA, OFF current 0.76nA. The simulation methodology is presented in Section IV.A.

### B. Characterization and Abstraction of Standard Cell

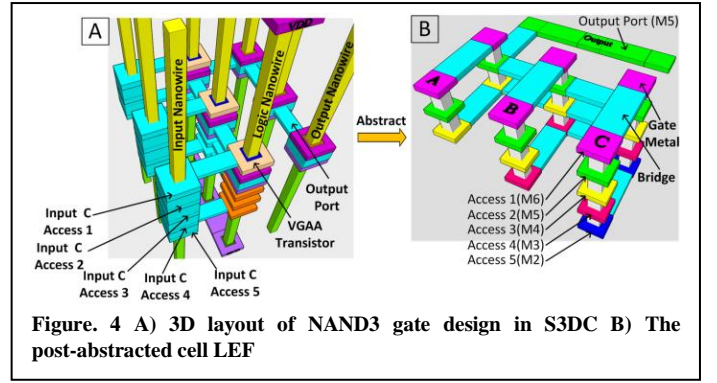

**Figure. 4 A) 3D layout of NAND3 gate design in S3DC B) The post-abstracted cell LEF**

Novel nanoelectronic devices do not have built-in models in traditional circuit simulators such as HSPICE. Therefore, device simulation data are used to create behavioral models for the n- and p-type VGAA junctionless devices compatible with HSPICE as explained in [19]. Then, the resistance and capacitance of interconnects in each standard cell are modeled using Predictive Technology Model (PTM) [20] and extracted from customized 3D layout, which is designed based on design rules of Skybridge fabric [14]. With the help of behavioral models and modeled interconnect RC values in each standard cell, we used Synopsys Siliconsmart to do power and timing characterization for each standard cell. Finally, these power and timing information of the all cells are merged into one cell library file (Lib file).

The position of intra-cell input/output pin access, the dimension of each cell and the used intra-cell routing metal are abstracted and written into the cell Library Exchange Format (LEF) [21] file that is used in Encounter based cell-to-cell routing. Fig. 4 shows the layout design and LEF abstraction of a 3D NAND3 gate. In the 3D layout the cell is designed with three rows of nanowires: input nanowires, logic nanowires and output nanowires. Each input nanowire with coaxial routing metal that connects one p-type transistor in pull-up network and pull-down network, acts as one input pin. The accesses to each input pin are vertically distributed and exactly aligned with the vertical grids (See Section II). This way, each pin port can be accessed both horizontally and vertically which expands the degree of accessibility and routability for the cell. The metal layer, position and dimension of each pin access are written into the LEF file. The pin accesses on output nanowire are designed and abstracted in a similar way. The metal pieces that are used for intra-cell routing are abstracted and defined as 'obstacle metal [21]' in LEF. The pin access, gate metal and bridge are abstracted with the same width and thickness (See Fig. 4 B).

### C. Imitation of Cell-to-cell Routing in Encounter

The commercial tool, Cadence Encounter, is originally designed for conventional 2D CMOS designs where the standard cells are placed and routed in a 2D array. The fundamental working mechanism is to place the cells with optimal half-perimeter Wirelength (HPWL) [22] of cell-to-cell interconnections, and then the cells are treated as 'black boxes' and routed based on their input/output pin position defined in the Cell LEF file. Similarly, this working mechanism can fit into our 3D routing because the S3DC's standard cells can also be envisioned placed in a 2D array. The input/output pin position of our 3D standard cells are abstracted and written into a Cell LEF file with the method discussed in last section. Then, Encounter places the cells and generates the optimal cell-to-cell routings based on the given Cell LEF file and post-synthesized gate netlist. And the cell-to-cell routings of S3DC are imitated with conventional routing components in Encounter; the bridges which form links between nanowires are imitated using routing metal which distributes vertically with different layers and routs along horizontal grids (See Fig. 5C); the coaxial routing nanowires which route along vertical nanowire are imitated using via stack. And each coaxial routing nanowire is placed in the cross center of two horizontal grids where Encounter would place via stack. This is set through the Cell LEF file. Since each via stack can only propagate one signal, we use the coaxial routing nanowire (See Fig. 1D) that has only one coaxial metal layer in our design.

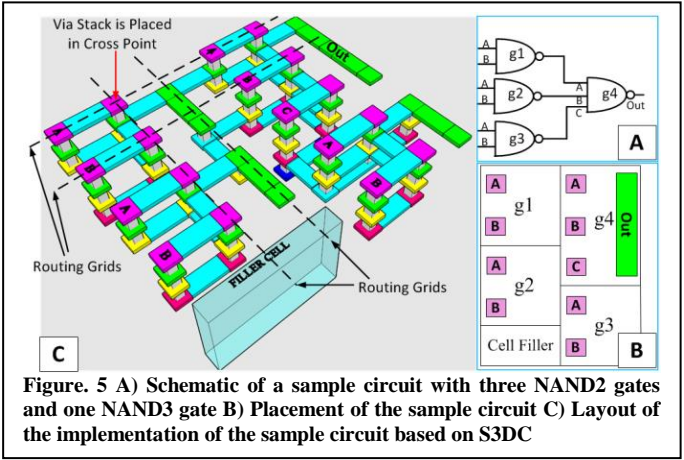

**Figure. 5 A)** Schematic of a sample circuit with three NAND2 gates and one NAND3 gate **B)** Placement of the sample circuit **C)** Layout of the implementation of the sample circuit based on S3DC

The feature sizes of routing metal (bridge) and via stack (coaxial routing nanowire) are designed by following the design rules of original Skybridge [14]. These rules are defined in the tech.lef file and .ict file. The .tch file, which sets the capacitance and resistance extraction rules, is generated using Cadence Techgen [23] with imported dimensions of metal layers (.ict file). The Cell LEF file, tech.lef file and .tch file are imported into Encounter as S3DC's fundamental design kit.

### D. Evaluation of Key Metrics

The key metrics are evaluated by using Synopsys Primetime with imported .spref file, Lib file and the synthesized netlist of the design. The .spef file contains the RC information of cell-to-cell routings which is extracted by Encounter. We perform Primetime statistical power analysis and timing analysis with the switching activity of both primary inputs and sequential outputs at 0.2.

Table I Comparison of Pin Access Number

| Pin Name | T-MI | S3DC | 2D CMOS |
|---|---|---|---|
| Pin A | 3 | 5 | 5 |
| Pin B | 2 | 5 | 6 |
| Pin C | 3 | 5 | 5 |
| Pin Output | 3 | 4 | 4 |

The area of the design is calculated by Encounter, and the die utilization ratio is set to be 0.6 which means 60% of the die area is used to place functional cells and the other 40% is used to place filler cells for providing extra routing space.

## IV. ROUTABILITY ANALYSIS

Conventional 3D ICs do not provide sufficient routability to address the high-density 3D routing which results in routing congestion issue in large-scale designs [6]. In this section, we carry out analysis to evaluate the routability of S3DC fabric and compare it with T-MI and 2D CMOS. DES [24], LDPC [24] and JPEG [24] cores are chosen as standard benchmark circuits to reflect the impact of routability on systems. We build these core designs in all technologies with a uniform technology node. The Nangate15nm PDK [25] is used in the designs of 2D CMOS and 3D T-MI CMOS, and the methodology in [6] is used for the benchmarking of 3D T-MI CMOS.

### A. Routing Congestion Issue in T-MI IC

As the most fine-grained monolithic 3D IC, the T-MI achieves 3D integration using standard cells in 3D style [6]. The 3D standard cell is designed with both reduced intra-cell RC and cell-to-cell interconnection which lead to better power efficiency compared with gate-level and block-level monolithic 3D ICs. The drawback is the 3D cells are designed with reduced access space in each input/output pin due to the shrunken footprint. Yet, the number of pins in 3D cell remains the same as 2D and die footprint is reduced about 50% [6], which means doubled routing density in T-MI's design. The heavily increased pin density and the reduced pin access space of standard cell leads to high routing demand and high-density routing.

### B. Improved Routing Scheme in S3DC

Since T-MI essentially mainly follows the conventional routing scheme in 2D CMOS, it has insufficient routability to meet the requirement of high-density 3D circuits. By contrast, the S3DC addresses cell-to-cell routing and cell pin access with a 3D mindset. Its standard cell is designed in vertically-composed 3D style (See Fig. 4A) where multiple transistors are vertically stacked on nanowires and the pin accesses to input/output ports are vertically distributed in different metal layers.

Compared with the conventional routing scheme in 2D CMOS, the S3DC's scheme has two main advantages: (i) the cell can be accessed in multiple metal layers which turns to reduce vertical routings, (ii) the access space of each pin is vertically expanded and multiple pin access are thus created in cell design. These two factors contribute to enhanced accessibility and routability for cells. The Table I shows the comparison of pin access number in 2D CMOS, T-MI and S3DC. The number of pin access in T-MI's NAND3 is nearly half of 2D CMOS's NAND3 while S3DC'S NAND3 is designed with even more pin accesses than 2D CMOS's.

### C. Routability Evaluation

The routability of 2D CMOS, T-MI and SB-CMOS are evaluated through analysis of routing congestion in their benchmark circuits. Generally, the routing congestion in IC design is caused by the high-demand or over-demand of routing resource [11]. Thus, routing demand is a key metric used to reflect routing congestion and evaluate the routing complexity for a design before detailed routing [27]. We carry out quantified evaluation using the relationship between the routing demand $l$ and the cell density $G$ per unit area [27]:

$$l \sim G^{r-0.5} (r > 0.5) \qquad (1)$$

Where G represents the effective number of cells that need to be routed in a unit square and r is a constant known as the Rent's exponent [26]. The value of G can be calculated using Rent's rule [26] as shown in equation (2). Rent's rule is an empirical observation about the relationship between the number of terminals (input/output pins) required by a design block to interface with its environment and the number of circuit components within the block [27]. It can be represented by the following equation:

$$E = A \cdot G^r => G = \left(\frac{E}{A}\right)^{\frac{1}{r}} \qquad (2)$$

where E is the number of terminals (input/output pins) in a unit square, A is the average number of terminals per cell. We assume all gates are distributed uniformly in the post-routed benchmark circuit. The parameter A is set to be 3 for each technology. The Rent's exponent r is set to 0.75 which is a typical value for large-scale designs [27]. Further, we use the pin number per micrometer square (pin density) as the parameter E. For 2D CMOS and T-MI, the pin density E is reported by Encounter after 2D placement for a certain design. For S3DC, the pin accesses of each cell are distributed in multiple metal layers but not limited in the M1 layer as T-MI and 2D CMOS. Therefore, we calculated the pin density of S3DC's design by the expression:

$$E_{S3DC} = \frac{\# \text{ of Pins}}{N \cdot S} = \frac{\# \text{ of Pins}}{S} \cdot \frac{1}{N} = E_{ENC} \cdot \frac{1}{N} \qquad (3)$$

N is the number of layers that are used to put pin accesses in our S3DC standard cell design. Its value is 5. S is the footprint of the die. $N \cdot S$ thus reflects the effective die area that is used to place cell pins. $E_{S3DC}$ denotes the real pin density in S3DC's design. $E_{ENC}$ is the pin density that is reported by Encounter that considers the cell pins distributed in a 2D plane and calculates the pin density using the die footprint S.

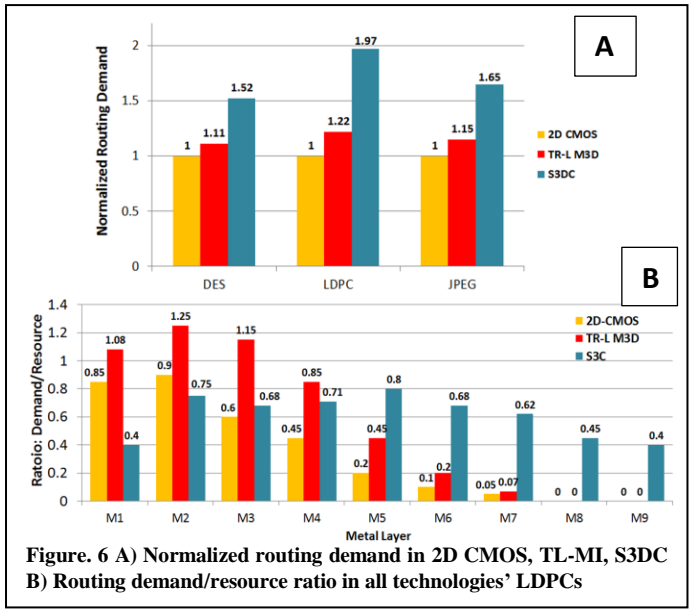

**Figure. 6 A) Normalized routing demand in 2D CMOS, TL-MI, S3DC B) Routing demand/resource ratio in all technologies' LDPCs**

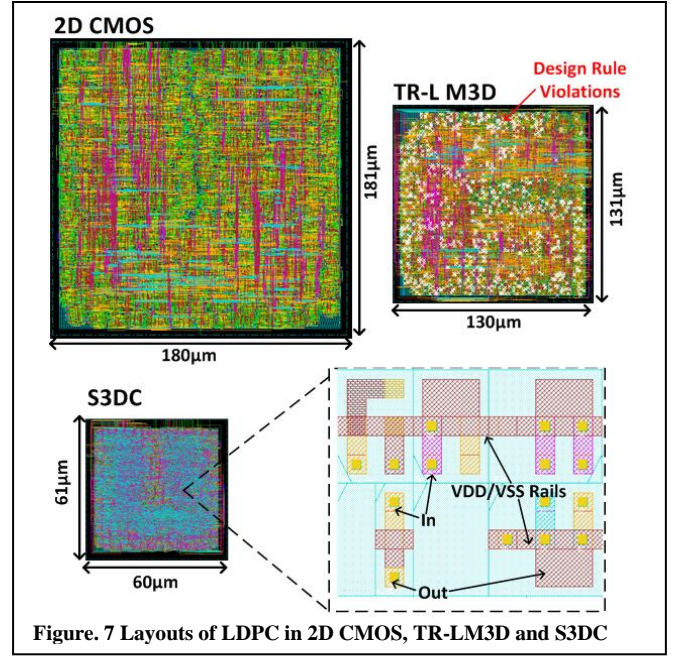

**Figure. 7 Layouts of LDPC in 2D CMOS, TR-LM3D and S3DC**

It can be seen that the S3DC's effective die area for placing pins is multiple of the die footprint since the pin accesses distribute in multiple metal layers while in 2D CMOS or T-MI's effective die area is equal to the footprint area. This contributes to significant pin density reduction in S3DC's designs in comparison to T-MI which in turn significantly reduces the routing demand.

Fig. 6A shows the normalized data of unit square's routing demand in each benchmark circuit for all technologies. The T-MI'S DES and JPEG designs have around 1.6x routing demand over 2D CMOS while S3DC's designs have up to 15% increased routing demand compared to 2D CMOS. For the interconnect dominated core, LDPC, the T-MI even shows 2x routing demand over 2D CMOS while S3DC has around 20% higher routing demand than 2D CMOS. It is observed that S3DC has slightly higher routing demand over 2D CMOS while it has up to 1.6x lower routing demand per unit square compared with T-MI.

Not only the high routing demand, the severe reduction of routing resource in 3D IC design is another important factor that results in over-demand of routing resource. The shrunken die area mainly leads to the routing resource reduction. Fig. 6B shows the ratios of routing demand and routing resource in LDPC core design for each technology. These ratios that represent different metal layers are all reported by Encounter after layer-by-layer detailed routing. It can be observed that the T-MI's LDPC design in encounter has over-demand routing in M1, M2 and M3 metal layers where the high-density routing for input/output pins are required. By contrast, for S3DC's LDPC design in Encounter, the routing demand distributes evenly in multiple metal layers with a maximum demand/resource ratio of 0.8 since the pin accesses of cells distribute in multiple metal layers.

Fig. 7 shows the layouts of LDPC core design in 2D CMOS, T-MI and S3DC, with clock tree, power delivery network, combination logic and sequential logic parts routed by Encounter. Due to high routing congestion rate, the T-MI's design is routed with thousands design rule violations. By contrast, the S3DC's design has 3x density over T-MI's design while it is routed without any design rule violation.

**Table II:** Results of Benchmarking

| Benchmark Name | Design Type | # of Cells | Best Frequency (GHz) | Total Wirelength (mm) | Wire Power (mW) | Cell Pin Power (mW) | Cell Internal Power (mW) | Total Power (mW) | Footprint | PPA |
|---|---|---|---|---|---|---|---|---|---|---|
| DES | 2D | 52380 | 4.6 | 99.00 | 0.32 | 1.12 | 2.52 | 3.96 | 1.00 | 1.00 |
| | TL-MI | 51450 | 5.3(+15%) | 71.28(-28%) | 0.26(-19%) | 0.95(-15%) | 2.09(-17%) | 3.30 (-17%) | 0.49(-51%) | 2.46 |
| | S3DC | 53450 | 4.1(-12%) | 30.69(-69%) | 0.15(-52%) | 0.19(-83%) | 0.91(-64%) | 1.25 (-66%) | 0.10(-90%) | 24.51 |
| LDPC | 2D | 36890 | 1.9 | 616.72 | 1.82 | 0.46 | 1.12 | 3.40 | 1.00 | 1.00 |
| | TL-MI | 34780 | 2.2(+17%) | 413.20(-33%) | 1.38(-24%) | 0.39(-16%) | 0.89(-19%) | 2.66(-22%) | 0.50(-50%) | 2.56 |
| | S3DC | 37689 | 1.7 (-10%) | 123.3(-80%) | 0.69(-62%) | 0.08(-81%) | 0.39(-65%) | 1.16(-62%) | 0.11(-89%) | 26.74 |
| JPEG | 2D | 297028 | 1.2 | 600.29 | 3.70 | 1.85 | 3.69 | 9.24 | 1.00 | 1.00 |
| | TL-MI | 287986 | 1.37(+14%) | 426.21(-29%) | 2.96(-20%) | 1.55(-16%) | 3.14(-16%) | 7.65(-20%) | 0.48(-52%) | 2.54 |
| | S3DC | 299076 | 1.1(-8%) | 180.08(-70%) | 1.85(-50%) | 0.33(-82%) | 1.29(-65%) | 3.47(-61%) | 0.11(-89%) | 24.57 |

Similarly, T-MI's DES and JPEG designs also have lots of design rules violations while S3DC's designs are routed with clear design rule check.

## V. BENCHMARKING RESULTS

The key metrics of the benchmark circuits are evaluated to reflect the design benefits contributed by routability. Further, a metric PPA which comprehensively involves power, performance and area (PPA), is used to evaluate the efficiency of technology. The metric PPA can be expressed as 'clock frequency/(power*footprint)'. The active power of each design is measured with uniform 1GHz clock frequency. And the area is reported by Encounter after placement. Table II shows evaluation results. The normalized footprint data shows that S3DC has up to 11x density against 2D CMOS, and the T-MI has around 2x density. The reduction of routing demand in conjunction with compact vertical 3D gate design contribute to about 3.3x shorter cell-to-cell wirelength which achieves up to 2.5x lower power against 2D while the T-MI only has up to 1.4x shorter wirelength and 1.25x wire power efficiency. Since the VGAA transistor has much lower parasitic capacitance than conventional Finfet with junction [15], our S3DC's standard cells have much lower driving capacitance, which achieves 6x lower cell pin power. The compact 3D standard cell design contributes up to 3x cell internal power efficiency. For interconnect dominated core, LDPC, the S3DC has 2.5x total power efficiency in comparison to 2D CMOS while the T-MI around 1.25x power efficiency. For the cell-dominated core, DES, the S3DC achieves up to 3x total power efficiency over 2D CMOS while the T-MI has 1.2x lower power compared to 2D. The S3DC yields impressive advances over T-MI in the PPA evaluation: S3DC shows up to 24.5x normalized PPA value against 2D CMOS while T-MI only have 2.5x PPA benefit against 2D CMOS. S3DC has around 10% performance degradation compared with 2D CMOS due to the usage of VGAA transistors, which have higher-resistivity channels [15]. This performance disadvantage however, can be overcome in multi-million transistor designs due to better routablity and shorter wire lengths [8].

## VI. CONCLUSION

In this paper, we study Skybridge-3D-CMOS (S3DC) fabric's routability and its impact on 3D IC designs. We investigate and compare with the state-of-the-art monolithic 3D IC, T-MI. A device-to-system design flow was developed to enable detailed evaluation. Compared with the T-MI, S3DC shows impressive routability for high-density routing in large-scale 3D designs which leads to 1.6x lower routing demand compared to TR-L 3MD. This routing demand reduction in conjunction with the compact vertical 3D gate design in S3DC contribute to significant benefits against 2D CMOS and T-MI: S3DC shows up to 3x power efieciency and 11x density against 2D CMOS while the T-MI has up to 1.25x power efficiency and 2x density compared to 2D CMOS. We expect that these benefits to improve in even larger systems and processors, paving a new path for 3D ICs.